\documentclass[10pt,a4paper]{article}
\usepackage{amsmath,amsfonts,amssymb}
\usepackage{mathrsfs}
\usepackage[body={17.5cm, 22cm},right=2.2cm]{geometry}

\newcommand{\ie}{{\textit{i.e.}~}}
\newcommand{\mm}{{\mathscr{M}}}
\newcommand{\ev}[1]{\langle #1 \rangle}
\newcommand{\cost}{{\mathrm{const.}}}

\newcommand{\IR}{{\mathrm{IR}}}
\newcommand{\vol}{{n}}
\newcommand{\vv}{{\mathscr{V}}}
\newcommand{\sem}{{\Delta}}
\newcommand{\stat}{{\mathrm{Stat}}}
\newcommand{\trial}{{\mathrm{Trial}}}
\newcommand{\Tr}{{\mathrm{Tr}}}
\newcommand{\etc}{{\mathrm{etc}}}
\newcommand{\D}{\mathcal{D}}

\begin{document}
\setcounter{page}{0}
\thispagestyle{empty}
\begin{center}

~

\vspace{5.cm}

{\Large
Gravity as an emergent phenomenon: a GFT perspective}\\[1cm]

{\large Lorenzo Sindoni \footnote{email: {\tt sindoni@aei.mpg.de}}}
\\[0.5cm]

{\small \textit{Max Planck Institut f\"ur Gravitationsphysik \\
Albert Einstein Institut\\
Am M\"uhlenberg 1, 14476 Golm, Germany
}}

\end{center}

\vspace{.8cm}
\vspace{0.3cm} {\small \noindent \textbf{Abstract} \\[0.3cm]
\noindent
While the idea of gravity as an emergent phenomenon is an intriguing one, little is known about concrete implementations that could lead to viable phenomenology, most of the obstructions being related to the intrinsic difficulties of formulating genuinely pregeometric theories.
In this paper we present a preliminary discussion of the impact of critical behavior of certain microscopic models for gravity, based on
group field theories, on the dynamics of the macroscopic regime. The continuum limit is examined in light of some scaling assumption, and the relevant consequences for
low energy effective theories are discussed, the role of universality, the corrections to scaling, the emergence of gravitational theories and the nature of their thermodynamical behavior.}

\vfill

\setcounter{footnote}{0}
\section{Introduction} \label{intro}
General relativity is often viewed as a low energy effective field theory \cite{burgess}, pretty much as a sort of ``Fermi-like-theory''
for geometrodynamics. This is largely due to its incompleteness at the classical level (due to singularity theorems), to power counting reasoning, suggesting that the Einstein--Hilbert action, with or without a cosmological constant, can be just the lowest order term of a series of terms, and, finally, to the fact that as a quantum field theory, the theory of an interacting
spin-2 massless particle suffers from the same problems of four-fermions interactions\footnote{On this, it is important to mention that, in recent years, this point has been reconsidered within the framework of asymptotic safety \cite{weinberg,asysafe}.}.

In the exploration of the physical content of general relativity, several clues have been collected
suggesting that spacetime and its geometry might be emergent concepts in the semiclassical
limit of an unknown theory, possibly pre-geometric in nature. On the other hand, all the 
quantum gravity models that we have at our disposal\footnote{With the exception of loop quantum gravity \cite{thiemann}, 
that takes very seriously the classical theory and proposes a Dirac quantization of it.}
are based not only on rewriting general relativity in terms of more suitable variables,
but typically imply a deep revision of the very concept of space and time \cite{danieleapproaches}.

It is often stated that the complete theory of quantum gravity will be able to cure the known diseases of perturbative quantization, as well as
to provide an explicit computation (at least in principle) of the effective couplings determining the corrections to
general relativity at increasingly high curvature\footnote{However, transplanckian features \cite{brandenberger} not necessarily related to planckian curvature regions might be difficult to grasp.}. Some results are known in this directions (let us mention, for instance,
corrections to supergravity from string theory and effective equations in loop quantum cosmology \cite{olmo}), but many other approaches,
most notably spinfoams \cite{perez,nicolai,alexandrov,dittrich} and group field theories (GFT) \cite{freidelgft,danielegftlibro,danielegft}, are still rather far away from the stage in which
concrete computations of physically relevant observables can be computed in a systematic way (see however, the
attempts in \cite{spinfoamcosmo, hellmann} within a cosmological context).

Many models have been proposed, but none has been proved decisive, to date, the reason of which can be traced back
to the difficulty of extracting quantitites that could be cast in the form of physical transition amplitudes or cross sections,
that can be concretely computed and compared with experimental observations.
A rather remarkable exception\footnote{The status of string theory in this respect is rather controversial, given that the contact
with effective field theories is more transparent, but still the correct limit including the standard model has not been identified.} comes from Causal Dynamical Triangulations (CDT) for which
many numerical results of direct interest are available. We refer to \cite{cdt1,cdt2,loll} for reviews, references and recent results. These last models, in particular, will be source of inspiration for many of the ideas
that we are describing in the following.

Parallely to these approaches, an increasing amount of work has been done in the direction of understanding gravitation as an emergent phenomenon. While the idea is not new, recently there has been a rather intense activity in the investigation of models based on condensed matter systems which display, in some regime, the emergence of effective metrics different from the one of the lab. On one side, the activity has gone in the direction of understanding better the properties of quantum fields in curved spacetime (Hawking effect, particle creation in cosmology and in strong fields, etc.) which is an important area that so far lacks experiments under controlled conditions. On the other side, there are some attempts to describe the properties of emergent dynamics for gravity, at least in very specific and limited situations.

All the models that have been described so far do make reference to some background metric structure, thus making their impact on quantum gravity research rather limited. Indeed, as a consequence of the background structures, Lorentz invariance and diffeomorphism invariance are often broken symmetries. This has an impact both at the kinematical level, with a naturalness problem accompanying the appearance of Lorentz symmetry at low energy \cite{iengo}, and at the dynamical level, with the graviton acquiring extra modes whose coupling to matter fields cannot be controlled in such a way to respect the observational constraints. To keep the idea alive, then, there are two possibilities. Either some specific mechanism is found to get rid of the problems encountered so far, or it is necessary to move forward and study completely different classes of models, where all the potentially dangerous features can be satisfactorily controlled, at least in principle.

Recently, the group field theory (GFT) approach \cite{freidelgft,danielegftlibro,danielegft} has raised a lot of attention as a possible realization of a (statistical mechanical) pregeometric
theory that allows a background independent formulation of the path integral approach to quantum gravity. In this sense, we believe that this class of models do represent the ideal arena for testing many of the ideas of emergent gravity \cite{hugeometrohydro,hucondensate,hamma1,hamma2,konopka1,konopka2,fotini,BEC,emergentsignature,humicromacro,wen,sigrav,verlinde,multibec}. One of the purposes of this paper is to elucidate the potentialities of such a framework in this perspective.

One of the main point of GFT is that semiclassical continuum limit should be found in correspondence to a suitably defined critical point of such a theory.
While many results have been accumulated so far (\cite{razvan1,razvan2,razvan3, rivasseau1,rivasseau2,freidelgurauoriti,dimare,baratinoriti, oritisindoni,baratingirellioriti,livineoritiryan,ryan,carrozzaoriti} just to mention some of them), still we lack explicit results concerning the presence of a critical point for the partition function in general models. However, in \cite{razvancritical} the appearance of critical behavior for a class of models (that we will be interested in, incidentally) have been
established, thus providing a strong basis for the discussion presented in the rest of this paper.

In this paper we would like to address the general idea of finding imprint, on low energy physics, of the continuum limit defined as a suitable phase transition. Of course, lacking a definite calculation describing the general critical properties of GFT we have to rely onto some assumptions, that are somehow arbitrary but reasonable enough to describe a large class of possible (and plausible) critical behaviors. 

The aim we have in mind is actually twofold. On one hand we will try to pinpoint some very specific effect, ideally the pattern of the modifications to the low energy action for gravity. On the
other hand, we will tackle the more general problem of the map between a microscopic pregeometric 
theory and a suitable effective field theory. While the results will be only partial, at this preliminary stage, they
do shed some new light on the topic. Indeed, while we cannot yet give an answer
to a physical question in terms of a number, we do propose a specific scheme to follow, in order to translate some formal
results of the underlying model into a concrete effective field theory. The choice of GFT is by no means a reduction
of generality: we believe that many of the comments that we will make can be exported to other approaches, even though
the very structure of GFT allows to address some issues in a very direct way.

Let us briefly sketch the scenario we are aiming at. The underlying pregeometric model (like GFT) will have the following properties:

\begin{enumerate}
\item the description of matter and gravity is given in terms of pregeometric structures (matrices, tensorial objects etc.);
\item the theory provides the partition function, as well as all the possible correlation functions of the theory;
\item the theory provides the Schwinger--Dyson equations and the Ward identities\footnote{In the rest of the paper we will refer to the Schwinger--Dyson equations implicitly including the Ward identities and all the possible
relations among the correlation functions that are implied by the properties and the symmetries of the partition function.} relating the correlation functions;
\item certain composite operators will lead to the summation\footnote{In general, we can expect that the sum is not really over the classical geometries etc., given that we know that this procedure would not really make sense. Rather, we might think that the sum is restricted to certain structures that somehow make the sum well defined (even just formally), for instance, simplicial complexes.} over all manifolds, geometries and field configurations having some specified boundary data, corresponding to the various transition amplitudes we could be interested in;

\item the microscopic theory depends on a number of free parameters, and the continuum limit is obtained by tuning the parameters to reach a critical point (hence a phase transition).

\end{enumerate}

In the language of critical phenomena, general relativity would be the result of a special
phase transition where the order parameter is an entire three-geometry (\ie a pair $(^{3}\mm,^{3}g)$) (or an entire four-geometry, depending on which formulation of the theory one wants to describe), while the Einstein--Hilbert action can be regarded as a Landau--Ginzburg action for the order parameter (for a presentation of these ideas, see \cite{amati,percaccicw,percaccihiggs}). 
Of course, this description holds whenever we are well inside the geometrical phase, where the fluctuations in the metric are negligible. Near the phase transition, approaching the pregeometric phase, one has to take into account more details, and the mean field description of gravity must be abandoned.

When trying to go for the statistical mechanical model underlying general relativity, one has to deal with the implementation of the symmetries and the structures characterizing the low energy effective theory (GR and standard model): diffeomorphism invariance, local Lorentz invariance, matter fields and internal gauge symmetries, matter couplings and the equivalence principle. While some results \cite{freideloritiryan,oritiryanscalar,girellilivineoriti,baratinoriti,baratingirellioriti,valentina} are already known in the literature, many questions remain open. For instance, one should clarify the emergence of diffeomorphism invariance in any dimension, the presence of a Lorentzian signature instead of a Riemannian one, the inclusion of appropriate matter fields and the recovery of effective field theories,the possibility of having anisotropic scaling between spatial and time directions (as in \cite{horava}), the clarification of the continum versus the semiclassical limit, the emergence of area laws for entropy associated to horizons and, obviously, unitarity of the quantum theory.

Concretely, in this paper we will start from the very simple situation of GFT describing pure gravity. Despite rather unphysical, this case already presents a number of interesting conceptual difficulties that must be addressed. The goal is to investigate the consequence of the presence of critical points for the partition function of gravity (as it is defined in this way), and in particular how it is possible to relate the critical behavior to a low energy/long range effective field theory, by examining the shape of the wavefunction and guessing the shape of the corresponding Wheeler--DeWitt equation. 

The results that we will obtain are essentially three. First of all, it is clear that the continuum limit is not giving automatically the semiclassical limit. Rather, the effective Hamiltonian operator that we will define will still contain all the quantum corrections (coming both from second quantization of the gravitational field and from third quantization effects). Second, we will draw a road map to pass from the knowledge of the critical behavior
of the microscopic model to the formulation of a long range effective field theory. 

At this level, to really discuss the equation of motion, we will have to deal with the precise form of the Schwinger--Dyson equations of the theory, and with the way in which they operate in the large volume limit. However, we will elaborate some strategies to effectively deal with the problem without necessarily passing from the microscopic theory (which is however a necessary step that can be postponed a bit, but not avoided). 

Despite this, we can still establish an expected but still important result: the emergent theory, \ie the coupling constants of the long range approximation will be determined by the critical behavior (in our case just by the critical exponent, but more complicated situations can be envisaged). Therefore, this leads to the conclusion that, close to criticality, the memory of the microscopic model and its parameters can be washed away, the only relevant feature being the universality class of the model.
In other words, universality seems to suggest that the microscopic parameters of the model cannot be immediately identified with
some macroscopic counterparts, as it is sometimes suggested, even when the construction of the microscopic model attempts to follow as closely as possible the macroscopic action that one hopes to obtain in the large scale limit.

In addition, we will also estimate the impact of the corrections to scaling (corrections that will contain the information distinguishing the different microscopic models having the same critical behavior)
when moving away from criticality and, finally, we will propose a concrete definition of entropy in terms of the microscopic models such that it might be compared with gravitational entropy as it is defined for semiclassical gravity (black hole entropy, De Sitter entropy, etc.).

Contextually, we get a third outcome from this analysis. 
What is rather surprising is that, despite taking as a case study a particular model for GFT that would seem to be completely blind to information about curvature (in a precise sense that we will specify later), it still seems that, on large scales, the effective dynamics can be non-trivial, at least in the cases we have considered in this paper. This might be a
coincidence, or, as it seems plausible, it is a manifestation of universality: given that we are looking at the system at criticality, the precise details of the
microscopic action might be washed out, and the final long range action (in an ideal world, general relativity) might be just the leftover at the critical point\footnote{We will expand on this in the conclusion, after all the evidences have been collected.}.

The plan of the paper is as follows.
In section \ref{models} we will introduce the setting, with the relevant definitions and the needed assumption. We will derive the general form of the thermodynamic potentials written in terms of the macroscopic extensive variables, like the total volume. After discussing some generalities of the problem of reconstructing the Wheeler--DeWitt equation in section \ref{wavefunctionwdw}, we will define, in section \ref{generating}, some generating functionals that
will be naturally associated to the presence of macroscopic boundaries, allowing us to define weighted sums over boundary geometries. Furthermore,
as a consequence of the scaling assumptions we will be able to argue the shape of the wavefunction in terms of the critical behavior of the system. 
Sections \ref{WDW2} and \ref{characteristiccurve} will be devoted to the reconstruction of the Wheeler--DeWitt operator necessary to develop an effective theory for
gravity, while in section \ref{corrections} we will give an estimate of the impact on the results of the corrections to scaling. After discussing some conceptual implications of these results on topics like the origin of the gravitational entropy (section \ref{entropy}), we will summarize the results in section \ref{conclusion}, where possible developments will be proposed. 
\section{The setting}\label{models}

In this paper we will take seriously the idea that spacetime is the result of a particular microscopic system being close to criticality.
We will go further with our assumptions and make an additional stipulation on some properties of criticality. The hypothesis we make is rather natural in light of what we know about critical phenomena. In particular, we will assume that the partition function for quantum gravity, thus defined, will display characteristic scaling behavior with respect to the parameters controlling the approach to the critical point. Of course, this should be accurately motivated only from the analysis of the underlying microscopic statistical mechanical model, but still
this represents a natural option, in light of the body of work related to matrix models \cite{ginspargmm}. The recent result \cite{razvancritical} strongly supports this point of view. However, the discussion presented in the rest of the paper could be adapted (at least in principle), to any other form
of critical behavior. 

\subsection{GFT: actions and notations}
Let us take, to make the discussion concrete, the case of the partition function of matrix models/GFTs. This gives (even if only formally) the sum over all the possible simplicial complexes (with appropriate topological restrictions), each one weighted by an amplitude which is uniquely determined by the original matrix model/GFT action. 

A group field theory \cite{freidelgft,danielegftlibro,danielegft} can be seen as a generalization of matrix models \cite{ginspargmm} and models like the ones proposed by Boulatov \cite{boulatov} and Ooguri \cite{ooguri} to the definition of the partition function of gravity in dimension three and four, respectively. The starting point is a field theory (for instance, complex scalar field theory) over a certain number of copies of the group manifold $G$,
\begin{equation}
\phi: \underbrace{G\times\cdots \times G}_{d \,times}\rightarrow \mathbb{C}
\end{equation}
where $d$ is the number of the dimensions of the spacetime that we want to discretize. In the remainder, to avoid clumsy mathematical formulae, we will use the notation
\begin{equation}
\phi_{a...b} \equiv \phi(h_{a},...,h_{b}).
\end{equation}
The theory is defined by means of an action
\begin{equation}
S_{GFT}[\phi] = \int (dh)^{d} \phi_{a_{1}...a_{d}} \bar{\phi}_{a_{1}...a_{d}} + \frac{h}{d!} \int (dg)^{K} V(\{h \})\phi_{a^{1}_{1}...a^{1}_{d}}...
\phi_{a^{d+1}_{1}...a^{d+1}_{d}} + cc.
\end{equation}
where $V$ is a certain function of the group elements encoding the combinatorics of the $d-$dimensional symplex.
Of course, more complicated choices are possible, enabling the implementation of specific properties of the
class of histories that one is summing over. These important modifications will not influence the general
gist of the discussion, and we will ignore them.
The partition function of the model is defined as usual by:
\begin{equation}
Z(g)= \int \D \phi \exp(-S_{GFT}[\phi]).
\end{equation}
It is possible to work also with the partition function defined with $\exp(-iS_{GFT})$. However, this would not add much to the discussion.
The very definition of such a partition function is far from obvious, and a suitable regularization procedure, defined in terms of a $1/N$ expansion \cite{razvan1,razvan2,razvan3}, must be adopted to provide a workable partition function.
Very briefly, the partition function will depend on the coupling constant, $g$, used to define the perturbative expansion of the theory. The Feynman graphs associated to this perturbative expansion are can be described by simplicial complexes, while the amplitude of each simplicial complex appearing in the sum is nothing else than a spinfoam amplitude. 
It is easy to see that, taking 
\begin{equation}
W(g)= \log Z(g),
\end{equation}
to eliminate the summation over disconnected graphs, we can define a partition function for $d-$dimensional quantum gravity\footnote{For a thorough discussion about the Feynman's path integral approach to quantum gravity, see \cite{hamberbook}}. We refer to the literature for more complete expositions.

The function $W$ can be given in terms of its perturbative expansion,
\begin{equation} 
W(g) = \sum_{\Gamma} g^{n(\Gamma)} A(\Gamma),
\label{eq:pertw}
\end{equation}
where the summation goes over all the simplicial complexes $\Gamma$ compatible with the Feynman rules of the model, $n(\Gamma)$ is the number of simplices of the complex and $A(\Gamma)$ is an amplitude that depends on the complex as well as on its colorings, when relevant, that can be understood as a Boltzmann weight factor in the statistical sum.

The function $W$ can be used to compute certain expectation values. For instance, it is easy to realize that, when using its perturbative expansion \eqref{eq:pertw}, the expectation value of the number of simplices reads
\begin{equation}
\langle n \rangle = \frac{1}{W} g\frac{\partial}{\partial g} W(g) = \frac{1}{W} \sum_{\Gamma}n(\Gamma) g^{n(\Gamma)} A(\Gamma),
\end{equation}
This quantity might be related, for instance, to an average volume, once the mapping between the combinatorial structures of the graphs and the metric information is established. Despite this is not obvious, we will do so, in the rest of the paper, as we shall specify in a moment.
 
Symmetry requirements on $\phi$ and on $V$ can lead to further specifications about the class of simplicial complexes that are entering the sum, as well
as their colorings with group representations and intertwiners which are encoding the geometrical properties (metric, connection, curvature) at the discrete level. Here we take a departure from the mainstream approach: while we take $V(\{h \})$ to be the correct one to lead to simplicial complexes of the appropriate dimensions, we do not use any projection on the field on the group manifold to enforce any form of invariance under special subgroups of $G\times... \times G$. As a consequence, it will be impossible to store any form of geometrical data on the simplicial complex, beyond the combinatorial structure, as it happens in more refined models implementing spinfoam amplitudes. The only thing we will be able to count is the number of simplices we are using. 

With respect to other GFTs for quantum gravity, like the one of Boulatov, which imposes a certain symmetry on the field $\phi$, necessary to reproduce the amplitudes of the Ponzano--Regge model for three dimensional gravity, we might think that we are indeed working with a theory of triples of edges that are not necessarily closing into triangles. However, we could also look at the model as a model for equilateral simplicial complexes. Besides the information on curvature encoded into the deficit angles (ultimately related to the combinatorial structure of the model), there is no other possibility to weight differently complexes with different curvature structure. This represents a great limitation imposed on our analysis, which will seriously affect our possibilities to make serious contact with models for quantum gravity proposed so far. We will come back on this point later, showing that this might be, in fact, a welcome property.

In analogy with matrix models, the continuum limit is thought to be achieved by tuning the coupling constant to a critical value, $g_{c}$, to be determined by further investigations.  The reason for this is that, at a critical point, the number $\langle n \rangle$ diverges, signalling that the ``average triangulation'' will contain an infinite number of simplices, a necessary condition to speak about a continuum limit. This is the reason why the understanding of the critical behavior is of paramount importance to understand completely the continuum, semiclassical limit of quantum gravity (in this particular approach).

For the rest
of the paper we will assume to have found this critical point.
Close to a phase transition, the free energy (as a function of the coupling constant) can be separated into a regular, analytic part and into a singular part. 
All the critical behavior is controlled by the singular part, while the regular part contains the information about the behavior of the system away from criticality.

\subsection{Criticality}
Let us start from the assumption that at criticality, the logarithm of the partition function for GFT has a singular behavior of the form:
\begin{equation}\label{eq:scaling1}
 W = a (g-g_c)^{\gamma}\left( 1+ f(g) \right)
\end{equation}
where $f(g)$ is a function of the coupling constant, analytic in a neighborhood of $g_{c}$. Of course, this is a rather
arbitrary assumption, to date, that consists in exporting verbatim the lessons of matrix models, at least in
their simplest incarnations \cite{ginspargmm}. Of course, one can imagine different forms for the non-analytic part
of $W$: only an explicit calculation can settle the matter. We will come back to the point in the concluding remarks. However, as said, the results of \cite{razvancritical} support this conjecture.

This stipulation \eqref{eq:scaling1}, which is the basic assumption for this section, amounts to a specification of the universality class of the critical point, determined essentially by the exponent $\gamma$. We assume that other contributions to
$W$ do not influence the critical behavior, while they represent the nonuniversal corrections away from criticality. These are precisely the kind of
corrections that could allow us to discriminate between two distinct microscopic models having the same critical behavior. We will neglect them, for now, and therefore
we will be describing only the general signatures of critical behavior, not the ones specific of the given model.

Generically, it is expected that the exponent $\gamma$ could depend on dimensionality of the model, the kind of interactions, presence of
additional degrees of freedom, pretty much in the same way in which it has been discussed in the context of matrix models.
Define, for convenience,
\begin{equation}
 \zeta = \log g.
\end{equation}
If we want to estimate the average volume (or surface, if we consider matrix model), we need to compute the average number of vertices\footnote{Again, remember that in the perturbative expansion, each vertex is dual to a $d-$simplex. Furthermore, the function $W$ can be expanded into
\begin{equation*}
 W(g) = \sum_{\Gamma: \mathrm{connected}} g^{n(\Gamma)} A(\Gamma)
\end{equation*}
where $n(\Gamma)$ denotes the number of simplices contained into the complex $\Gamma$, and $A(\Gamma)$ is an amplitude
determined by the combinatorial structure of the Feynman graph (including symmetry factors).
}, which is seen to be:
\begin{equation}
\ev{n_{v}}(g) = \vol =\frac{1}{W} g\frac{\partial W}{\partial g} = \frac{1}{W} \frac{\partial W}{\partial \zeta},
\end{equation}
This number is the average number of elementary cells in the Feynman expansion.
Clearly, $g\rightarrow g_{c}$ corresponds to the limit of infinite
combinatorial volume\footnote{Sometimes it is mentioned that the GFT coupling constant
is related to the cosmological constant. Indeed, as these expressions show, it is in fact
related to a variable that is the thermodynamical dual of the average volume.}. Notice that with one parameter at our disposal
we cannot really probe more than this, at this stage. As usual, to decide whether this number is physical or not
we need to see whether this is somehow related to the physical volume probed by matter fields. To uncover other properties
of the average configuration, like topology, curvature etc., we would need more external
parameters/coupling constants.

In our case, as a consequence of \eqref{eq:scaling1}
\begin{equation}
 \vol = \frac{\gamma g}{g-g_c} = \frac{\gamma e^{\zeta}}{e^{\zeta}-e^{\zeta_c}},
\end{equation}
where we are dropping the angular brackets to avoid clutter.
We can invert these relations, to get:
\begin{equation}
 g = \frac{g_c n}{n-\gamma}, \qquad
\zeta= \log\left(\frac{g_c n}{n-\gamma} \right)
\end{equation}
Notice that these relations holds only close to criticality, when $\vol \gg 1$.

This number is just combinatorial in nature, and does not immediately lead to a physical volume, for which the metric structure needs to be established. To do this we need to introduce a length scale in the
game, $\epsilon$ and define:
\begin{equation}
V = \epsilon^{d} \vol.
\end{equation}

The nature of $\epsilon$ requires further discussion. If we consider just $g\rightarrow g_{c}$ we just get  an infinite $\vol$ limit,
not a continuum limit. As it is well known, in the case of lattice field theory, taking an infinite number of lattice points will not lead to any continuum limit of the theory.
To speak of the continuum limit we need to check that the correlation length of the field theory becomes macroscopic with respect to the lattice spacing.
Therefore, to discuss in a physically relevant way the continuum limit we need to introduce matter fields, representing rods and clocks which will
give the precise mapping between the combinatorial quantities defined at the level of the Feynman expansion and the dimensional quantities of physical interest. In the setting we are working, this issue is left aside and it is implicitly assumed that all this procedure of introducing matter fields etc. is
directly included in the appearance of the parameter $\epsilon$ which translates the combinatorial volume (counting of elementary volumes) into a physical
volume. We could imagine, for the sake of the discussion, that $\epsilon$ is the ratio between the correlation length  $\xi$ (defined in combinatorial terms),
and the lattice spacing $a$:
\begin{equation}
\epsilon = \frac{a}{\xi}.
\end{equation}
This ratio will represent the ratio between the Planck scale and the ruler constructed with the correlation length of the given field.
With this in mind, we will drop the $\epsilon$ from our expressions.

\subsection{Consequences}

Define the free energy to be:
\begin{equation}
 F = -\log W 
\end{equation}
so that $W$ is maximized when $F$ is minimized. In our case
\begin{equation}
 F = - \gamma \log(g-g_c) + \cost
\end{equation}
This is a thermodynamic potential of the system when $\zeta$ is used as an intensive macroscopic
variable to describe its properties. On the other hand, we could define another thermodynamic
potential, say $\Omega$, defined as the Legendre transform\footnote{With the conventions
commonly used in thermodynamics to define the procedure.} of $F$, where now the system
is described by the conjugate macroscopic variable, $\vol$:
\begin{equation}
 \Omega(n) = F(\zeta(n))- \zeta(n) n = -\gamma \log \left(\frac{\gamma g_c}{n-\gamma}\right) - n \log\left(\frac{g_c n}{n-\gamma} \right)
\end{equation}
Let us manipulate this expression
\begin{equation}
 \Omega(n) = -
\gamma \log ({\gamma g_c})+\gamma\log(n-\gamma)
- 
n \log(g_c) 
-n \log \left(\frac{ n}{n-\gamma} \right)
\end{equation}
Now, as we said, we are interested into the limit in which $n \gg 1$, for which:
\begin{equation}
 n-\gamma \approx n, \qquad \frac{n}{n-\gamma} \approx 1
\end{equation}
Here it is clear how $\gamma$ represents a sort of elementary volume. One would be
tempted to use it to generate and expansion of $\Omega$ in powers of $\gamma$. However
for these corrections to be really important, we need $n\approx \gamma$, which is a regime
we are excluding from our assumption of being near criticality. If we were considering
a system which is too far away from criticality, we would need to include all the
nonuniversal contribution to the partition function that we have neglected, here,
from the very beginning, which presumably will give relevant contributions. We will describe them in
section \ref{corrections}. 
For the moment, we will neglect all the terms of order $1/n$.
In this limit:
\begin{equation}
 \Omega(n) \approx \gamma\log(n) - n \log(g_c)
\end{equation}
where we have neglected a constant which does not depend on the thermodynamic variable $n$.

We can use $\Omega(n)$ to define a probability measure for the volume $\vol$,
\begin{equation}
P(\vol) = \frac{1}{\Xi} \exp(-\Omega(n)) = \frac{1}{\Xi} n^{\gamma} \exp(-\alpha n),
\end{equation}
where 
\begin{equation}
 \alpha = -\log{g_c},
\end{equation}
and $\gamma$ is the critical exponent.

The normalization function $\Xi$ is easily computed
\begin{equation}
 \Xi = \int_{n_{\IR}}^{\infty} \vol^{\gamma} \exp(-\alpha \vol) ,
\end{equation}
where $n_{\IR}$ is an infrared regulator needed to get rid of the small volume contribution,
that is not within the domain of applicability for this model.
Keeping this in mind, we will consider the case $n_{\IR} = 0$, for convenience in the
calculation but also because, for the case $\gamma >0$, the term $n^{\gamma}$ automatically suppresses
the contribution coming from small volumes.

Therefore
\begin{equation}
 \Xi = \int_{0}^{\infty} \vol^{\gamma} \exp(-\alpha \vol)d\vol = \int_{0}^{\infty} du e^{-u} \frac{u^{\gamma}}{\alpha^{\gamma}} \frac{du}{\alpha} = \frac{\Gamma(\gamma+1)}{\alpha^{\gamma+1}} 
\end{equation}

With this statistical measure we can compute average value for various quantities, like the average
value of some power of the volume
\begin{equation}
 \langle \vol^{k}\rangle = \frac{1}{\Xi} \int_0^{\infty} \vol^{\gamma+k} \exp(-\alpha \vol) d\vol = \frac{1}{\alpha^k}\frac{\Gamma(\gamma+k+1)}{\Gamma(k+1)} 
\end{equation}
This expression can be further simplified using the recursive properties of the Euler's gamma function
\begin{equation}
 \Gamma(z+1) = z\Gamma(z)
,\qquad
 \Gamma(z+k) = \left(\prod_{i=0}^{k} (z+i) \right) \Gamma(z)
\end{equation}
to give
\begin{equation}
 \langle \vol^{k} \rangle = \alpha^{-k}\left(\prod_{i=1}^{k} (\gamma+i) \right) 
\end{equation}
Define the average volume to be
\begin{equation}
 \vv = \langle \vol \rangle = \frac{\gamma+1}{\alpha}.
\end{equation}
Clearly, for consistency
\begin{equation}\label{eq:mean}
 \vv \gg 1 \Leftrightarrow \alpha \ll 1
\end{equation}
which means that $g_c$ must be very close to one, so that the most probably a weak coupling expansion
cannot be used at all. This is consistent with the fact that, to have arbitrarily large Feynman diagrams
we cannot truncate the perturbative expansion at a given finite order.

Similarly, we can define the spread of the distribution, by considering 
\begin{equation}\label{eq:spread}
 \sem = \frac{\langle \vol^2\rangle - \vv^2}{\vv^2} = \frac{1}{\gamma+1}
\end{equation}
Equations \eqref{eq:mean} and \eqref{eq:spread} suggest that we can replace $\alpha$ and $\gamma$,
which are defined by means of the microscopic theory, with some corresponding macroscopic quantities
which control the mean value and the typical fluctuations around the mean value.
In particular:
\begin{equation}
 \gamma = \frac{1}{\sem} - 1
,\qquad
 \alpha = \frac{1}{\sem \vv}
\end{equation}
whence
\begin{equation}
 P(\vol) =  \frac{\Gamma(\sem^{-1})}{\alpha^{\sem^{-1}}} n^{\sem^{-1}-1} \exp\left(-\frac{1}{\sem} \frac{\vol}{\vv}\right)
\end{equation}
The way in which $\sem$ enters this expression suggest to think about it as a ``temperature''.
Nonetheless, this identification is only formal. Interestingly, $\sem$ is determined by the critical exponent $\gamma$
which describes the way in which the continuum limit is approached, and hence cannot be varied ad libitum
without varying the (universality class of the) underlying microscopic model.
Introducing the notation
\begin{equation}
 \beta= \frac{1}{\sem},
\end{equation}
our expression for $P$ becomes:
\begin{equation} 
 P(\vol) = \Gamma(\beta) \left(\frac{\vv}{\beta}\right)^{\beta} n^{\beta-1} \exp(-\beta \vol/\vv)
 \label{eq:distribution}
\end{equation}
Notice that the requirement that the resulting distribution has a very small spread around the mean value
(which in light of our interpretation of this kind of models corresponds to a spacetime which is essentially
classical), $\Delta \ll 1$, which implies that $\gamma \gg 1$. In terms of the ``inverse temperature''
$\beta$, this condition can be seen as the requirement that the ensemble is kept at a very small temperature (but again, this interpretation is only formal, since the critical exponent $\gamma$ is not a parameter which can be easily controlled, being
determined by the underlying dynamics of the model).

\section{Wavefunction and the reconstruction of Wheeler--DeWitt} \label{wavefunctionwdw}

The discussion of the previous section was focused on a very primitive approach. While for models
of two dimensional quantum gravity the use of equilateral triangulations proves to be rich enough to get some
nontrivial results (see, for instance \cite{ginspargmm}), for higher dimensional structures we have the additional
problem of introducing, in one way or the other, a notion of curvature at the microscopic level. For instance, in GFT the gravitational action
can be introduced by certain projections on the group manifold that one is considering, in such a way that, when 
developing the perturbative expansion of the theory, each vertex of the Feynman graphs is associated to a suitable amplitude which
encodes, in the appropriate language, a discretized version of the Einstein--Hilbert action. For instance, in the
case of three dimensions, the vertex is associated to a Wigner's $6j$ symbol, which, as proved by Ponzano and
Regge \cite{ponzanoregge} is deeply related to the cosine of the action used in the Regge calculus approach
to classical (and quantum) gravity \cite{williamsreggecalculus}. 

In the case of causal dynamical triangulations for four dimensional quantum gravity, besides the volume term counting the
number of 4-simplices,
the curvature is introduced by appropriately weighting the 2-simplices on which curvature is localized (in a distributional)
sense. There, one has two distinct parameters, one for the curvature and one for the volume, to control
the system, to drive it to criticality and to obtain different phases. We refer again to the literature \cite{cdt1, cdt2}
for more details and references. 

In the case of models like GFT, the only parameter at our disposal is the GFT coupling constant, with the interaction term
encoding the entire gravitational action. Therefore, it seems difficult to disentangle the various contributions from
curvature and cosmological constant already at the level of the partition function. Therefore we need something different. An option is to consider certain correlation functions of the GFT model which might have a more direct physical interpretation for quantum gravity.
In this perspective, there are two natural candidates: the ``wavefunction'' for quantum geometrodynamics and the transition amplitudes between different boundary geometries.

With the term wavefunction we will denote the particular correlation functions of the GFT model corresponding to the summation of triangulations possessing
a single boundary with given data, representing a discrete geometry. The transition amplitudes are those particular correlation
functions that correspond to the summation over triangulations possessing two boundaries with (in principle) two different sets
of data encoding geometrical properties. Examples of transition amplitudes have been considered in the past, especially within
the realm of the spinfoam approach to quantum gravity, and lately to quantum cosmology \cite{spinfoamcosmo}. 
On the other hand, the wavefunction approach originated already with DeWitt \cite{dewitt}, in an attempt to define canonical quantization
of gravity based on the implementation of the constraints of general relativity as functional differential operator acting on a 
function defined over the configuration space of canonical general relativity, superspace. We will follow this second point of view,
in particular in light of the Hartle--Hawking proposal \cite{hartlehawking} that fits pretty naturally our framework.

The Hartle--Hawking proposal is a prescription to define the wavefunction of the universe in terms of a Euclidean path integral
defined over compact geometries having a unique boundary whose intrinsic geometry is the particular point of the superspace where the
wavefunction is evaluated. Clearly, our choice to work with
the partition function defined with $\exp(-S)$ instead of $\exp(iS)$ fits particularly well this scheme.

There are interpretational issues on what we should define as wavefunction. In ordinary quantum mechanics, a wavefunction
is usually defined as a projection of the quantum state on an orthonormal basis constructed out of some complete set of commuting Hermitian
operators, associated to certain functions of the classical phase space of the theory. In the case of gravity the situation is more
complicated, given that the natural domain of definition of a wavefunction, superspace, needs to be put in correspondence
with operators whose eigenstates are, indeed, 3-geometries. Even if we would succeed in providing such a construction, there
would be the difficulty of relating these states with the ones that we could typically construct by means of discrete structures. In this respect, the most refined and rigorous results are obtained within the formalism of
spin networks.

Again, let us take the case of GFT and the wavefunctions that we can construct out of them. Let us focus on the four dimensional case.
The quantities that we can construct (and perhaps compute within some approximation scheme) are some correlation functions
corresponding to three dimensional simplicial complexes, with given topological properties, colored with some representation label, say $j$ if we consider $SU(2)$.
Let us take the point of view that these quantities are really possible states of quantum gravity, solving some discretized version
of the Wheeler--DeWitt equation. Let us denote them as $|\Delta, \{j\}\rangle$. Clearly, what we would need to define is a
scalar product\footnote{Or a suitable projection.} between these states and (hypothetical) states associated to smooth three geometries $|(^{3}\mm,^{3}g)\rangle$,
\begin{equation}
\langle (^{3}\mm,^{3}g)| \Delta, \{j\}\rangle
\end{equation}
which would allow to make a precise connection between the discreet and the continuum formalism. This is not just a side issue. Take the
case of a single line of length $L$: we could discretize it with $n_{1}$ segments of length $L/n_{1}$ or with $n_{2}$ segments of
length $L/n_{2}$. Both these discretizations are legitimate, a priori, and hence they might contribute, with different weight, to the definition
of the state that we call ``a macroscopic loop with length $L$''. The problem of the coarse graining of microscopic geometrical information
encoded into graphs, and the definition of states which have definite global geometrical properties is still subject of active research \cite{oritipereirasindoni}.

Furthermore, as we have already said, to be really able to address physical questions,
one has to introduce matter fields, that can be used to introduce notions like physical time (as it happens, for instance, in 
the case of quantum cosmology \cite{kieferqg}). The realization of group field theories containing realistic matter fields in their Feynman expansion is still work in progress (see, however, \cite{dittrich}).
In addition, we have to include the notion of continuum limit/criticality, and, finally, of semiclassical limit, which will involve the
definition of suitable quantities and parameters associated to these procedures.

While this constitutes a whole research program that cannot be exhausted in one paper, we can still try to make conjecture and work out
some of the consequences of them, pretty much within the lines of what we have discussed in the previous section. In particular, we will
implement, in a different way, the scaling hypothesis that we have made in the previous section, now treating the three geometry as a thermodynamical
variable. 
Let us be more specific on this. While the complete specification of a geometry might be a hopeless task, we might imagine that we can treat situations with high symmetry so that many degrees of freedom can be effectively neglected. Therefore, we assume to restrict the analysis to
describe situations in which we can completely identify the geometry that we are interested in by means of a certain number of functionals,
for instance the total volume, the mean curvature etc. For the sake of the argument let us assume that we can restrict this infinite family of functionals (local and nonlocal)
to a few of them, say the total volume and the average scalar and square curvature 
\begin{equation}
f_{1}[q] = \int_{\mm}\sqrt{q} ,\qquad f_{2}[q] =\int_{\mm} \sqrt{q}R(q) ,\qquad f_{3}[q]=\int_{\mm} \sqrt{q} R^{2}(q)
\end{equation}
where we are fixing once for all a background manifold $\mm$, and we are restricting $q$ to be of a certain family, such that these three coordinates give
a complete specification of which element of the family of metrics $q$ we are talking about. This approach has all the limitations
of the symmetry reduced systems, but still will give us the possibility of concretely draw a road map.

Our wavefunction will be constructed out of these quantities
\begin{equation}
\psi[q]=\Phi(f_{1}(	q),f_{2}(q),f_{3}(q))
\label{eq:wf}
\end{equation}
and, in principle, its shape can be computed on the basis of suitable computations made
at the microscopic level \cite{oritisindonihh}. For now, we will assume that this has been done, and that the wavefunction \eqref{eq:wf} is known explicitly as a function.  
By the very construction, this quantity obey the diffeomorphism constraint equations:
\begin{equation}
\nabla_{i} \left(\frac{\delta}{\delta q_{ij}} \psi[q]\right) =0,
\end{equation}
given that $\psi$ is constructed with diffeomorphism invariant quantities. The true question is: what is the Hamiltonian constraint such that this wavefunction
solves the corresponding Wheeler--DeWitt equation? As a subordinated question, what is the semiclassical limit of the theory obtained in this way, and finally, what are the quantum corrections that we can expect, with their associated phenomenology?
Such a question is nontrivial even in more traditional approaches to the path integral for quantum gravity \cite{halliwell,hartlehalliwell}.

The situation is clear: a formalism like GFT, or spinfoams, promises to compute quantities like $\psi[q]$, or more general matrix elements. However, to really make contact with the language of effective field theories, which is the language in which we can talk about experimental tests of theories, these days, we need to translate this quantity in terms of Lagrangian, or, equivalently, differential equations.

As we said, the theory that allows us to compute the wavefunction should also provide some relations between the
wavefunction computed at different points. This is the point at which the Schwinger--Dyson equations for the correlation
functions come into the game. These equations will be, generically, recursion relations between correlation functions
with different number of points and with different colorings. When performing the continuum limit, one has not only
to compute the continuum limit of the correlation functions, but also the continuum limit of the equations relating them.
This is what has been done, within the context of matrix models, with loop equations \cite{ginspargmm,dijkgraaf,migdalloop}. Loop equations are recursive
equations relating matrix correlation functions associated to loops with different lengths, and, ideally, in the continuum
limit, they should give rise to the equations for the continuum dynamics \cite{loops1,loops2,loops3}. 

For the case at hand, the situation is still not clear. It is plausible that, given the relation between the constraints of GR and diffeomorphisms, one should use the Ward identities associated to the symmetries realizing,
at the microscopic level, diffeomorphisms. To this purpose, the work in \cite{baratingirellioriti} is of special relevance. Nonetheless, still much more is needed, especially in higher dimensions.

Without the Schwinger--Dyson equations, or some recurrence relations among correlation functions, the formulation
of an effective field theory becomes rather difficult and subject to uncontrolled arbitrariness. In this paper we will able
to generically address this arbitrariness, but a solution can be obtained only when the exact statistical properties of GFT
will be elucidated.

In the remainder, we will try to extract as much as possible from the only knowledge of the critical behavior, without having at our disposal any information about the dynamics in the critical limit. While this represents a limitation, we will still be able to uncover a number of interesting properties.

\section{Generating functions}\label{generating}

A key tool to define consistently the wavefunction, within a combinatorial approach, is by means of a suitable generating function. This
will be done by adapting the procedure used for matrix models to the case of higher dimensional spaces. In the case of (single) matrix models, it is customary to define the generating functions
\begin{equation}
L(x) = \Tr\left \langle \frac{1}{x\mathbb{I}-M} \right\rangle_{c}, \qquad L(x,y) =\left\langle \Tr \left(\frac{1}{x\mathbb{I}-M} \right) \Tr\left( \frac{1}{y
\mathbb{I}-M} \right) \right\rangle_{c}, \qquad \etc
\end{equation}
where $\langle  ... \rangle_{c}$ denotes the expectation value obtained counting only the connected diagrams, $x,y$ are real numbers and $M$ is the matrix for which the statistical system is defined. 
These particular correlation functions represents summation over two dimensional triangulations with single or multiple boundaries, given by loops whose
average length is controlled by the ``fugacities''  $x,y$, representing the thermodynamically conjugated variables to the
length of the loops. In other words,
\begin{equation}
\frac{\partial L(x)}{\partial x} = \langle \ell \rangle
\end{equation}
where $\ell$ is the length of the loop. This result might be derived in the same way in which we have counted the average number of simplices from the derivative of the partition function $W$.
Furthermore, these correlation functions, as a consequence of Schwinger--Dyson equations, obey a hierarchy of equations \cite{ginspargmm}, known as loop equations,
which represent a direct path to the understanding of the continuum limit of the models \cite{migdalloop}.

The reason why this approach should be preferred is that it explicitly addresses the question about what superpositions of different microscopic configurations (\ie boundary lengths) will contribute to the state macroscopically labeled by a certain geometrical property (\ie the macroscopic length). Clearly, different choices of generating functions will generate sums that will
weigh  in different ways the various contributions, representing different statistical ensembles, according to the number of parameters introduced.

In this section we propose a generalization of these generating functions to the case of higher dimensional models. The objective is to define in a consistent and controlled way the sum over different discretizations of the same boundary.
For our purpose, we will need to consider just the case of a single boundary, but we will envisage a way to define the generating function for multiple boundaries. Let us start with the general structure of the result that we want to obtain. We want to define in a consistent way a weighted sum of correlation functions which have specific combinatorics associated to a specific combinatoric structure for a certain D-1 manifold, something of the form
\begin{equation}
G(\lambda) = \sum_{\Gamma_{D-1}} \lambda^{n(\Gamma_{D-1})} G(\Gamma_{D-1})
\end{equation}
where $\lambda$ is a parameter needed to control the average size of the boundary (\ie a chemical potential dual its volume) and
$G(\Gamma)$ a GFT correlation function associated to the boundary triangulation $\Gamma_{D-1}$.

The crucial difficulty in defining a sum over all the possible discretizations is to generate them. However, for this purpose we could use the fact that with an auxiliary GFT associated to a $D-1$ dimensional model we can accomplish this task\footnote{An analogous idea, applied to a rather different aspect of GFT has first been
suggested by F. Hellman \cite{franknote}.}. 
Let us work, for instance, in three dimensions. Then, define
\begin{equation*}
G(\lambda, g) = \int \D \psi \D\phi \exp \left[-\left( \frac{1}{2}\int (dg)^{2} \psi_{ab} \psi_{ba} + \frac{\lambda}{3} \int (dg)^{3} \psi_{ab}\psi_{bc} \psi_{ca} \phi_{h_{1}h_{2}h_{3}} \right)+
\right.
\end{equation*}
\begin{equation}
\left.
+
\left(
\frac{1}{2}\int (dg)^{3} \phi_{abc} \phi_{cba} + \frac{g}{4} \int  \phi_{abc}\phi_{aef}\phi_{dec}\phi_{dbf}  \right)\right]
\label{eq:generatingfunction}
\end{equation}
where $h_{1}=g_{a}g_{b}, h_{2} = g_{b}g_{c}, h_{3} = g_{c}g_{a}$.
The first term in the exponential generates a perturbative in $\lambda$ which would consists of all the possible two dimensional triangulations, if
instead of the $\phi_{abc}$ we would have just $1$. The presence of $\phi_{abc}$ implies that the Feynman expansion will be in fact a sum of correlation functions for the three dimensional model, \ie a sum of amplitudes of random 3D triangulations with random boundaries.

By modifying the properties of the auxiliary group field theory (presence of colors, different combinatorics, modified kinetic and potential terms), we can control the class of boundaries we can sum over. The study of this generating function goes beyond the scope of the paper. However, it
represents a new direction for the study of these kind of models of quantum gravity and deserves further study both for its definition and for the investigation of its properties.

Clearly, the definition given in three dimensions can be exported to the case of higher dimensions. Also, by inserting other auxiliary group field theories, we can define generating functions which are associated to summation over simplicial complexes with multiple boundaries.

The definition \eqref{eq:generatingfunction} provides a summation over three dimensional simplicial complexes with arbitrary two dimensional boundaries,
and the parameter $\lambda$ represents the thermodynamical dual to the size of the boundary (the only variable that at the moment we are able to control).

The usefulness of this generating function is clear: it allows us to define consistently the sum over all boundary triangulations, once we specify the statistical property of the sum, controlled by $\lambda$ and $g$. In particular, we can conjecture the existence of a critical behavior that corresponds to a continuum limit in which both the boundary and the bulk are triangulated with an infinite number of blocks (the caveat on the definition of the continuum limit in terms of
appropriate correlation lengths, discussed in section II, still applies). As such, we can imagine that, near the criticality, the connected part of $G$,
\begin{equation}
Q(\lambda, g)= -\log G(\lambda,g)
\end{equation}
could have a behavior similar to the one of the free energy W studied in section II, namely
\begin{equation}
Q(\lambda, g) \approx (\lambda-\lambda_{c}(g))^{\delta} (g-g_{c}(\lambda))^{\gamma},
\end{equation}
were we have included the possibility that the critical values $\lambda_{c},g_{c}$ (and, in principle, also the critical exponents) have a dependence on the value of the other coupling constant. For the moment, we will drop this possibility, but we will come back to it in section \ref{corrections}.

As a consequence, we can proceed in perfect analogy with section \ref{models} and get to the same results, once we replace $g$ with $\lambda$.
In particular, we can infer that a boundary with size $A$ will have a probability
\begin{equation}
P(A) = \frac{A^{\delta}}{\Theta}  \exp(-\alpha' A), \qquad \Theta = \int_{0}^{\infty} A^{\delta} \exp(-\alpha')A dA
\label{eq:scaling2}
\end{equation}
where 
\begin{equation}
 \alpha' = -\log{\lambda_c}
\end{equation}
Clearly, this formula is valid only in the continuum limit and in the case in which we are very close to the critical point $(\lambda_{c},g_{c})$, and neglects
a lot of possibilities for the critical behavior. In this respect, a direct calculation cannot be circumvented. However, given that to date we do not
have it, we will try to get as much as we can from this conjecture.
In the rest of the paper we will try to extract as many consequences we can from equation \eqref{eq:scaling2}.
We will come back later on the issue of corrections to scaling.

\section{Reconstruction of WDW}\label{WDW2}

Now that we have collected some ideas concerning what kind of behavior we should expect, let us come back to the problem of translating a wavefunctional in superspace in terms of a Wheeler--DeWitt equation.
One of the interesting question raised by the previous analysis is: what is the effective (field) theory underlying the results obtained via GFT?
In the investigation of (causal) dynamical triangulation, the issue of the formulation of an effective theory describing the various phases of
the model is of great importance. There, the presence of a microscopic partition function that allows a more or less direct connection with a
transfer matrix or an Hamiltonian allows us to go as far as the construction of an effective fiducial Hamiltonian, at least in
certain simplified cases \cite{benedettilollzamponi}.

We assume that, due to the nature of spacetime as a continuum limit, the microscopic system is at criticality, and
that this fact manifests itself into the scaling properties of the correlation functions, in terms of the macroscopic variables:
\begin{equation}
\psi[q] = (f_{1}[q])^{\gamma} \exp(-\beta (f_{1}(q))^{\alpha})
\end{equation}
as an example. Assume, for the moment, that we further reduce our
superspace to the space of spheres, and that the only free parameter determining totally the geometry is the radius of the
sphere, $r$.
Then:
\begin{equation}
\psi[q] = (r)^{3\gamma} \exp\left(-\frac{\beta}{\hbar} r^{3\alpha}\right)
\end{equation}
where we have changed slightly the notation to include the information that what does matter is the action of the system,
weighted in units\footnote{The emergence of units is indeed a big foundational problem that needs to be addressed.} of $\hbar$.
Furthermore, we have introduced $3\alpha$ to keep in mind that, in $3+1$ dimensions, when $\alpha = 1$ we recover the
standard large volume behavior of the wavefunction.
We can now try to make contact with what we know. We want to find a linear differential operator $\hat{L}$, such that
\begin{equation}
\hat{L} \psi(r) = 0.
\end{equation}
Moreover, we know the semiclassical limit, the equation, treated in the WKB approximation, should lead to an eikonal equation
which is nothing else than the Hamilton--Jacobi equation for the classical limit, that is general relativity. As a consequence,
we can cook up a linear (self-adjoint) differential operator with at most two partial derivatives with respect to $r$ 
\begin{equation}
\hat{L}_{\trial} = \hbar^{2} \frac{\partial^{2}}{\partial r^{2}} + i\hbar u(r) \frac{\partial}{\partial r} + \left(v(r)-\frac{i}{2} \frac{du}{dr}(r)\right)
\end{equation}
and with a number of
free functions to be determined.

In this case
\begin{equation}
\frac{\partial}{\partial r} \psi = \left( \frac{3\gamma}{r} - \frac{3\alpha \beta}{\hbar} r^{3\alpha-1} \right) \psi
\end{equation}
whence:
\begin{equation}
\frac{\partial^{2}}{\partial r^{2}} \psi 
=\left[ \frac{3\gamma(3\gamma-1)}{r^{2}} +\frac{9\alpha^{2} \beta^{2}}{\hbar^{2}} r^{6\alpha-2} -\frac{C}{\hbar} r^{3\alpha-2}  \right] \psi
\end{equation}
Now, if the wavefunction is real (as it happens for ground state functions of one dimensional systems), and in this parametrization this means $\beta \in \mathbb{R}$, the function $u(r)$ is identically zero while the function 
$v(r)$ is simply
\begin{equation}
v(r)=-\left[{9\alpha^{2} \beta^{2}} r^{6\alpha-2} - C \hbar r^{3\alpha-2} +\frac{3\gamma(3\gamma-1)}{r^{2}}\hbar^{2}\right]
\end{equation}
Obviously, the theory will not be truncated in an expansion in $\hbar$, given that the starting point is not a semiclassical limit, but rather a large volume
continuum limit in which all the quantum corrections are sistematically included.  In this sense, this shows how the critical behavior encodes just the continuum limit, and not the semiclassical one, which should be achieved otherwise.
Let us stress again that $\hbar$ in
itself needs to be defined, as the reference action, and that the models, being purely combinatorial, need to be manipulated in order to identify, at criticality,
the emergence of scales and physical units. For a similar case, involving the emergence of the speed of light, see  \cite{emergentsignature}.

The operator found in this way can be used easily to develop a systematic effective field theory and to make definite predictions about physical
observables, rather than using the wavefunction itself\footnote{On this point, see the comment in the introduction of chapter 4 of \cite{feynmanhibbs}.}. 
In principle, knowing the effective potential in the WDW equation we might want to go backwards and define a classical effective action leading to it in the suitable limit (perhaps involving some effective order reduction, as in \cite{simon,sotiriou}).

Extrapolating the discussion from the mere Riemannian quantum gravity models, it is clear that in the way in which we have proceeded, scaling is limited to the dependence on spatial properties, and the theory emerging needs not to be the 3+1 version of a Lorentz invariant theory (especially when matter fields are included). Consequently there must be a key argument to keep Lorentz symmetry under control. This should be provided by the Schwinger--Dyson equations and all the other relations among the correlation functions, which are enforcing at each level the microscopic symmetries (for instance, diffeomorphism invariance of the macroscopic theory as encoded in terms of a symmetry of the microscopic GFT). 

The fact that Lorentz invariance must be implemented already at the microscopic level seems
to be rather compelling in light of the evidences that have been collected in the latest years
about effective field theories with Lorentz violation effects systematically included \cite{liberatimaccione}. However, the realization of the
symmetry at large scales is by no means guaranteed, since there might be some form of
spontaneous symmetry breaking which cannot be addressed without the complete understanding
of the continuum limit.

\section{Characteristic curve method}\label{characteristiccurve}
There is another way to get a better grasp of the inverse problem of finding the Wheeler--DeWitt equation given a solution, and it uses the method
developed to solve first order partial differential equations, the characteristic curve method. The key tool we would like to exploit is the relation between
the theory of linear partial differential equation, the WKB method and the Hamilton--Jacobi formulation of classical mechanics.

The idea is simple. If the dynamical equation of motion has the form of a differential equation (of arbitrary order), it is possible to find approximate solutions by means of the eikonal approximation, which consists of approximating the solution with a function of the form $\exp(iS(x)/L)$, where $L$ is a small parameter (in appropriate units). While the method is used extensively in quantum mechanics, specialized to the case of second order PDE, its validity is more general \cite{WKB}.

Within this framework, given the Hamilton function, we can find the Hamiton--Jacobi equation it satisfies, trying to go backwards to uncover the nature of the differential equation which is ultimately its origin. The method, as we will see, will not be free of ambiguities, but still they will be drastically reduced, with respect to the very hand waving analysis of the previous section.

As it is well known \cite{couranthilbert2}, to solve a first order PDE of the form
\begin{equation}
H(\partial S, x) = 0
\end{equation}
where $x \in U \subset \mathbb{R}^{n}, S:U\rightarrow \mathbb{R}$, one can proceed by defining
\begin{equation}
p_{i} = \partial_{i} S
\end{equation}
and hence the canonical system
\begin{equation}
\dot{x} = \frac{\partial H}{ \partial p}
\qquad
\dot{p} = -\frac{\partial H}{\partial x}
\qquad
\dot{S} = p \dot{x}
\end{equation}
With this method it is possible to generate solutions of the PDE obeying the desired initial conditions.
Now, we are interested into the problem of finding $H$, once $S$ is given. For instance, 
\begin{equation}
S(x)= f(x)
\end{equation}
with a certain, given function $f(x)$.
Knowing $S$ we can compute $p(x) = f'(x)$
and hence the task is reduced to the problem of finding a function $H$ such that
\begin{equation}
H(f'(x),x) = 0
\end{equation}
Clearly, there will be more than one function satisfying this condition, and we will be left with this fundamental ambiguity
that needs to be solved with other means.

Instead of giving some abstract general derivation, we will specialize to the cases in which we are interested, that is to say, to functions of the form%
\begin{equation}
f(x) = \log( x^{\gamma}\exp(x^{b})) =x^{b} - \gamma \log x
\end{equation}
which is the sort of functions we could be interested in, with $x$ being related to some form of average size.
Clearly:
\begin{equation}
p(x) = bx^{b-1} - \frac{\gamma}{x}
\end{equation}
On the other hand:
\begin{equation}
\dot{p} = \frac{\partial p}{\partial x} \dot{x} = -\frac{\partial H}{\partial x}
\end{equation}
whence we obtain a PDE for the function $H(p,x)$:
\begin{equation}
\frac{\partial p}{\partial x} \frac{\partial H}{\partial p}+ \frac{\partial H}{\partial x} =0 
\end{equation}
or, using the expression for $p$:
\begin{equation}
(b(b-1) x^{b-2} +\frac{\gamma}{x^{2}}) \frac{\partial H}{\partial p} +  \frac{\partial H}{\partial x} =0
\end{equation}
This PDE can be solved via the characteristic curve method. In particular, we can define a new canonical system:
\begin{equation}
x' = \frac{d x}{du} = 1, \qquad p' = \frac{dp}{du} = b(b-1) x^{b-2} +\frac{\gamma}{x^{2}}
\end{equation}
It is straightforward to realize that the function $H$ is constant along the curves:
\begin{equation}
p - b x^{b-1} + \frac{\gamma }{x} = \cost
\end{equation}
which means that:
\begin{equation}
H(x,p) = \Phi \left(p - b x^{b-1} + \frac{\gamma }{x}\right),
\end{equation}
with $\Phi$ an arbitrary function. To fix $\Phi$, we still have to use the information coming from the SD equations, a step that cannot be completely avoided. However, the ambiguity in the system is entirely contained in the unknown function $\Phi$, and hence, by making appropriate assumptions, one could develop a systematic effective theory for the classical limit.

We can exploit this technique, then, to find the classical Hamiltonian which is associated to the WKB approximation of the solution of the equation we do not need to solve (having already its soution). In our context, the WKB parameter will be the inverse of the elementary volume.

It is clear that with this method we can, at least in very simplified cases, make contact between
the results of certain quantum gravity models concerning certain correlation functions and a language that
is more close to the one of effective field theory. Clearly, what has been shown here is just what happens
in the case of a minisuperspace model, in which all the complexity is reduced to the dynamics of a finite number of
degrees of freedom.

The hope is that the method can be generalized to the case of functional calculus (see, for instance, the detailed derivation
of Einstein's equations by means of the Hamilton--Jacobi method in \cite{gerlachhj}), and that it can be used, at least in principle,
to develop a systematic effective field theory, for instance in the case of small perturbations from homogeneous configurations.
To achieve this, however, we need more than what we have showed here.

\section{Corrections to the critical behavior}\label{corrections}

In this section we are going to discuss briefly the issue of the non-universal corrections to the critical behavior and their impact on the
kind of observables that we have just mentioned. Again, as our case study, we will investigate the corrections to critical behavior of
the form:
\begin{equation}
W(g) \approx (g-g_{c})^{\gamma} \left( 1 + b_{1} (g-g_{c}) + b_{2}(g-g_{c})^{2} + O((g-g_{c})^{3}) \right)
\end{equation}
where $b_{1},b_{2}$ are the non-universal coefficients (\ie model dependent) that control the first departure from the scaling behavior.

\subsection{Corrections to the partition function}

As in section \ref{models}, define
\begin{equation}
n= \frac{1}{W} g \frac{\partial W}{\partial g}.
\end{equation}
In this case we get
\begin{equation} 
n = \frac{\gamma g_{c}}{(g-g_{c})} + (\gamma+b_{1} g_{c}) + (b_{1}-b_{1}^{2}g_{c} + 2 b_{2}g_{c}) (g-g_{c}) + O(g-g_{c})^{2}
\end{equation}

Notice that the presence of a non-vanishing $b_{1}$ leads already to a correction to what would be the first term obtained expanding the results
of section \ref{models} beyond the lowest order. Working at the given approximation is it possible to find that:
\begin{equation}
(g-g_{c}) = \frac{a_{1}}{n} + \frac{a_{2}}{n^{2}} + \frac{a_{3}}{n^{3}} + O(n^{-4}),
\end{equation}
where
\begin{equation}
a_{1} = \gamma g_{c}; \qquad a_{2} = \gamma g_{c} (\gamma+b_{1} g_{c}); \qquad a_{3} = \gamma g_{c} [\gamma^{2} + b_{1}^{2} g_{c}^{2} +
b_{1} \gamma g_{c} (3-b_{1}g_{c})]
\end{equation}
As a consequence, quantities like $\Omega(n)$ become, to the next to the next to the leading order,
\begin{equation}
\Omega(n) = -\log g_{c} n - \gamma \log n + \left( -\frac{a_{1}}{g_{c}} + \gamma \log a_{1} \right) + \frac{c_{1}}{n}
\end{equation}
where $c_{1}$ is a calculable coefficient. Hence, taking the exponential,
\begin{equation}
P(\Omega) = n^{\gamma} \exp \left( - \alpha n - \frac{c_{1}}{n} \right) 
\end{equation}

In principle, one should insert this expression back into the machinery used to get the effective equations of motion. However, at large volume the effect of the logarithmic corrections due to the scaling (\ie the prefactor $n^{\gamma}$) are still larger than the non-universal corrections. On one side this is an interesting result, since it means that we could in principle detect as a first deviation from GR some operators that are associated purely to the critical behavior of the underlying model, while the corrections due to the granular nature of spacetime might become relevant at higher energies. 
On the other hand, this result might be shifting even farther away the access to the microphysics of the system, which is of course disappointing
for the quest of the identification of the microscopic theory describing spacetime.

\subsection{Corrections to the wavefunction}
The corrections to the partition function are not enough to understand what is the real impact of the critical behavior on low energy effective theory.
In the case of amplitudes computed with the given auxiliary fields, however, we need to include more into the game. Indeed, as anticipated in section
\ref{generating}, we should expect not only corrections due to the fact that the system is not exactly at criticality, but also the fact that the critical exponents
might in turn depend on the position in the space $(\lambda,g)$.

It is convenient to define
\begin{equation}
\epsilon = g - g_{c}^{0}; \qquad \eta = \lambda-\lambda_{c}^{0},
\end{equation}
where $ (\lambda_{c}^{0},g_{c}^{0})$
is the critical point we are approaching\footnote{Notice that, due to the presence of an auxiliary group field theory, $g_{c}^{0}$ might be different from the critical point for the partition function.}. In the neighborhood of this point we might Taylor-expand the various functions that are involved. For instance
\begin{eqnarray}
\lambda_{c}(g) \approx  \lambda_{c}^{0} + d \epsilon + O(\epsilon)^{2}, & g_{c}(\lambda) \approx g_{c}^{0} + c \eta \\
\gamma(\lambda) \approx \gamma_{0} + a \eta + O(\eta^{2}), & \delta (g) \approx \delta_{0} + b \epsilon + O(\epsilon^{2})
\end{eqnarray}
whence
\begin{equation}
Q(\lambda,g) = (\lambda-\lambda_{c}(g))^{\delta(g)} (g-g_{c}(\lambda))^{\gamma(\lambda)}(1 + m_{1} \eta + n_{1} \epsilon + O(\epsilon^{2},\eta^{2},\epsilon\eta) )
\end{equation}
might be approximated by
\begin{equation}
Q(\lambda,g) \approx (\eta-d\epsilon)^{\delta_{0}} (\epsilon-c\eta)^{\gamma_{0}} \left( 1+m_{1} \eta+n_{1} \epsilon + 
b\epsilon \log(\eta-d\epsilon) + a \eta \log(\epsilon-c\eta) + O(\epsilon^{2},\eta^{2},\epsilon\eta)\right)
\end{equation}
The running of the critical exponents will lead to logarithmic terms which give rise to additional corrections to the wavefunction. 
Unfortunately we will not be able to work out an explicit analytical calculation which takes these terms into account.
However, what is crucial to observe here is that the presence of such corrections will lead to an inevitable mixing between 
$\epsilon$ and $\eta$ in the definitions of the bulk volume $n_{bulk}$ and the boundary volumes $n_{boundary}$:
\begin{equation}
n_{bulk} = \frac{1}{Q} g \frac{\partial Q}{\partial g}; \qquad n_{boundary} = \frac{1}{Q}  \lambda \frac{\partial Q}{ \partial \lambda},
\end{equation}
and hence, when going for a representation of the probability amplitude in terms of these quantities, there will be contributions to
the wavefunction which will take into account not only the continuum limit on the boundary, but also the continuum limit in the bulk, the corrections to scaling and the corrections to the critical exponents. 
Still, the logarithmic term coming from the prefactor will dominate over the corrections due to granularity, at long range.

\section{Dynamics of gravity and gravitational entropy}\label{entropy}

One of the most intriguing aspects of GR that requires an explanation is the nature of the 
relationship between gravity and thermodynamics \cite{frolov}. This relationship seems to be crucially
related to the presence of horizons, in the broad sense of null surfaces, and the fact that
general relativity can be rewritten in terms of laws describing the evolution of them.

Clearly, this requires an explanation. Using the results of section \ref{models} we now see how it is possible to provide a definition
of entropy, by introducing a statistical distribution that describes the property of an ensemble of microscopic
states, weighted by the thermodynamical potential $\Omega(\vol)$. 

On purely statistical grounds we can define a notion of entropy associated to this distribution function,
\begin{equation}
 S_{\stat} = - \int_0 ^{\infty} P(\vol)\log(P(\vol))d\vol 
\end{equation}
To compute this expression we make use of standard manipulations:
\begin{equation}
 \int_0 ^{\infty} x^{a} e^{-x} \log(x^a e^{-x}) dx = \left. \left[\frac{d}{d\eta} \int_{0}^{\infty} \left( x^a \exp(-x) \right)^{\eta}\right] \right|_{\eta=0} 
\end{equation}
We get:
\begin{equation}
 S_{\stat} = \log\Xi -(1+\gamma+\gamma\log(\alpha)) + \gamma \frac{\Gamma'(\gamma+1)}{\Gamma(\gamma+1)} =
\end{equation}
\begin{equation}\label{eq:statisticalentropy}
 S_{\stat} = \log \Gamma(\beta) - \beta \log(\vv\beta^{-1}) -(\beta-1)  
\log(\vv \beta^-1) + \Phi(\beta) = -(2 \beta+1) \log(\vv) + \psi(\beta)
\end{equation}
where $\psi(\beta)$ is a function of $\beta$ alone. Remember that $\beta$ controls the fluctuations: the larger
is $\beta$, the smaller are the fluctuations. Furthermore, $\beta$ is controlled by the exponent $\gamma$ that describes
the approach to criticality, and hence is a universal property of the system. This suggests to isolate all the
dependence on the term $\psi(\beta)$. 

The natural comparison that we can make is between the result \eqref{eq:statisticalentropy} and the partition function
for euclidean quantum gravity with a cosmological constant, computed with the method of gravitational instantons in
\cite{gibbonshawkingdesitter}.
This comparison shows a basic discrepancy: we do not get the term scaling with the area of the cosmological horizon.
This in a certain sense obvious, given that in this simple model we have not introduced any mean to control the contribution
to each amplitude of the appropriate discretization of the Ricci scalar. The mismatch between the starting microscopic
actions might be the first reason. A second reason could be that the critical behavior generating spacetime as we know
it is of a different type. Unfortunately, knowing $S_{\stat}$, which makes reference only on the average properties of the
statistical distribution, does not allow us to go backwards and reconstruct the critical behavior of the partition function.
While this term is per se intriguing (opening up a direct connection between macroscopic calculable properties of spacetime to microscopic ones associated to criticality), it is probably
not the whole contribution to the gravitational entropy: it might be only the entropy associated with
the uncertainty in the total volume. 

In fact, there is another statistical entropy that we can define, just observing the way in which the
partition function is constructed in terms of triangulations $\Gamma$
\begin{equation}
W = \sum_{\Gamma} g^{n(\Gamma)} A(\Gamma)
\end{equation}
The function $g^{n(\Gamma)}A(\Gamma)$ can be seen as a distribution function in the space of triangulations, 
and as such we can compute its statistical/Shannon entropy:
\begin{equation}
S_{W} = -\log{W} + \frac{1}{W} \sum_{\Gamma} g^{n(\Gamma)} A(\Gamma) \log \left( g^{n(\Gamma)} A(\Gamma) \right) =
-\log W + \vol + \sum_{\Gamma} g^{n(\Gamma)} A(\Gamma) \log (A(\Gamma))
\end{equation}
This entropy would be much more compelling in light of the intuition that spacetime itself is a thermodynamical
notion, result of the averaging of a lot of microscopic configurations.

Unfortunately, we cannot yet devise, at the level of the microscopic action like the GFT one, a way to introduce a dummy parameter
such that we can obtain this quantity by taking, for instance, a derivative of a generating function to get the desired result
(as it happens, for instance, in the case of the evaluation of the entropy in the canonical ensemble as a suitable derivative
with respect to the inverse temperature).
What is clear from this expression, however, is that this entropy will be a certain function of $g-g_{c}$, or, equivalently, of $\vol$
which is an extensive function of the average triangulation, in principle scaling with its volume.

In fact, the main lesson that we can get out of the analysis
of thermodynamics of gravity is that the thermodynamical properties are not localized on given regions,
like in the case of black holes, but rather each infinitesimal region of spacetime, in its own, is a thermodynamical system \cite{jacobson}.
The case of black holes is just a situation in which it is particularly easy to uncover the relationship.

However, functions like the entropies that we have defined so far depend on volumes, rather than surfaces, and tend
to refer to global properties of the statistical system (\ie of the classical spacetime) rather than to local properties as we should have to match the behavior of local Rindler horizons.

There are at least three possibilities. First of all, the kind of formalism we have in mind is tailored for the case of Euclidean
spaces, rather than Lorentzian spaces. Signature must be appropriately introduced, but also the dynamics probably needs
to be revised, to get the appropriate scaling of entropies with areas and not with volumes. In this sense, the Lorentzian nature would be an essential building block.

The second possibility is that horizon entropy is a sort of ``red herring'': after all, it has been proven that horizon thermodynamics
is not at all limited to the case of general relativity, and that many different geometrical theories of modified (but Lorentzian) gravity can be
manipulated in such a way to reproduce some thermodynamical behavior, at the price of introducing non-equilibrium terms \cite{elingguedensjacobson,brusteinhadad,chircoliberati,chircoelingliberati}.
What is really important, in the pregeometric case, is that the amplitudes computed (and hence their semiclassical limit) match with
some form of gravitational theory.

The third case is in a sense mirror symmetric to the second one, and it is related to the fact that the kind of entropies that we are
trying to define have their own dignity, but their definition is only formal, in the sense that they do not enter the balance
of physical entropy encoded in the generalized second law (where all the forms of entropy, including the gravitational one,
do enter the balance).

To settle this issue, one needs to introduce matter fields explicitly and hence define gravitational entropy in such a way that, together
with matter entropy defined in the conventional way, it is possible to formulate consistently thermodynamics (with a generalized second law).
This is a point of view dual to the one advocating the need to introduce matter fields for the very definition of the physical geometry.

In this respect, there is
another important (and related) issue involves the formulation of a low energy effective field theory for gravity: the implementation of local Lorentz invariance. Given the present constraints on
Lorentz-violating effects in particle physics \cite{liberatimaccione}, we are forced, if we do not want to introduce some additional
microscopic symmetry yet to be discovered \cite{barceloliberativisser, liberatinaturalness, mattingly}, Lorentz invariance at the fundamental level. Similar arguments have been shown to hold, in simplified cases of emergent gravity, for the equivalence principle \cite{multibec}.
However, one might still have some Lorentz breaking effects (spontaneous breaking as in \cite{kosteleckysamuel}, or anisotropic
scaling features like in \cite{horava}).

There is some interesting phenomenological implication that deserves to be studied.
Within formalisms like GFT, the wavefunction and the correlation functions, even in the continuum limit, cannot be expected
to implement locality at the level of spacetime, given that the starting point is nonlocal in an auxiliary space, and that spacetime emerges only in a statistical sense. Therefore, 
the effective action will be, generically, a complicated nonlocal object. The limit of local quantum field theory will have to emerge in some regime, with locality as implemented in the standard model being only a long range phenomenon.
In this respect, it will be interesting to compare these models to other models of emergent locality discussed in other contexts of emergent/analogue
gravity \cite{hamma1,hamma2,konopka1,konopka2,fotini,wroclaw}, 
as well as to develop a full phenomenological framework to test nonlocality effects in particle physics processes, by constructing effective field theories with nonlocality features.

\section{Concluding remarks and outlook} \label{conclusion}

In this paper we have tried to address the issue of the continuum semiclassical limit of class of models 
for quantum gravity like group field theory, basing the analysis on the conjecture that near criticality
the behavior of the connected part of the partition function is essentially that of a homogeneous function. 
Despite this restriction, that is due to criteria of simplicity rather than on a full fledged calculation which is
still lacking, the general logic can be exported to more general case. 

The most significant step in the investigation is the introduction of suitable generating function for amplitudes
or wavefunctions, in section \ref{generating}, that allow us to use the critical behavior of the statistical system to infer the
shape of the wavefunction of the universe in the sense of Hartle and Hawking, and hence, by techniques discussed in section \ref{WDW2} and \ref{characteristiccurve}, used to
draw very primitive continuum effective field theories, at least in the case of minisuperspace models.

Clearly, much more work needs to be done, but still the results we are collecting here seem to be promising, showing
for instance that it is possible that the first corrections to the Einstein equations might tell us something about the continuum
limit (its universality class, in particular), rather than the corrections due to the detailed (and largely unknown) microscopic
structure of spacetime. This is of course much more interesting than the opposite situation, since it would allow us to
discriminate between different microscopic models for spacetime just looking at their universality class.

The procedure elucidates also the origin of the macroscopic coupling constants coming from the microscopic theory.
While it is still premature to envisage a systematic identification of the various terms in a gravitational action in terms
of curvature invariants, the general intuition is clear. They will be the outcome of the critical properties of the system
used to define the continuum limit, with corrections coming from the granularity of spacetime (encoded in the non-universal terms).
This is particularly significant since it sheds a different light on the problem of the renormalization of the gravitational
coupling constants. For instance, the actual value of the cosmological constant will not be determined by any vacuum
energy counting: there is no such a thing as an energy, in the pregeometric formalism we are using. The only thing that
does matter is the way in which the continuum limit is approached, as well as the underlying dynamics. This is perfectly
analogous to what has been discussed in \cite{finazziliberatisindoni}(see also \cite{alexander1,alexander2}). In light of this, the cosmological constant problem
must be translated into a question regarding the critical behavior of the underlying theory. A pregeometric approach
to the cosmological constant problem seems to be a rather intriguing one, deserving further investigations.

While all these topics do have their relevance, a calculation that can elucidate many of the previous issues would
be the determination of the statistical properties of De Sitter spacetime.
De Sitter spacetime shares many thermodynamical properties with black hole spacetimes, given that cosmological horizons
do define thermodynamical quantities associated to their dynamics \cite{gibbonshawkingdesitter}. However, the nature of these thermodynamical
properties is much more interesting than the corresponding ones for black holes. Indeed, while one can imagine
that the thermodynamical properties of black holes can be inferred by the properties of the degrees of freedom
more or less localized within the black hole event horizon, the same cannot be said for De Sitter space. Indeed,
cosmological horizons are observer dependent, given that the space is maximally symmetric. Hence, if the thermodynamical
behavior of DeSitter space can be associated to some pregeometric degrees of freedom, they must be ``equally distributed''
in spacetime: in this sense, probing De Sitter spacetime we should be able to have more direct access to some of
the properties of the fundamental degrees of freedom of spacetime \cite{maldacenastrominger,balasubramanian}.

As we have pointed out at the end of section \ref{models}, we do not have yet any positive result in this direction.
In a scenario in which only equilateral simplices are glued (via a GFT-like formalism), it is impossible
to include information about curvature, and hence to introduce even a very simple form of gravitational action.
We need to go beyond this model.

As we have shown, by introducing certain auxiliary fields we can work with amplitudes and wavefunctions, and thus
infer from there the properties of the effective dynamics by a comparison with. However, we still lack
a systematic way to control the boundary data in a local way: while it is easy to ask questions concerning
homogenous spatial sections, it is unclear how to introduce concretely the inhomogeneities (and the parameters controlling them)
associated to gravitons.

Nonetheless, this result suggests that at least part of the logarithmic corrections to the entropy of de Sitter space
can be originated by the statistical fluctuations of the system near criticality, and hence lead to
the conclusion that these corrections, if detected, can give direct access to the microscopic theory underlying classical spacetime, 
shedding some light onto the continuum limit.
However, a reproduction of the De Sitter entropy calculation would not be enough. Indeed, what makes gravitational
entropy relevant is that it enters the balance of entropy flows, and hence a big deal would be to check that
the entropy as it would be defined is really counted in a generalized second law.

To conclude, let us examine briefly the other important limit that we have neglected so far, the semiclassical limit. From
the results of sections \ref{WDW2} and \ref{characteristiccurve} it is clear that, whenever we translate the results
of the critical behavior of GFT in terms of a continuum field theory, it will still be quantum field theory \footnote{Again, working in
the Euclidean sector with the partition function does not change the result.}, containing all the quantum corrections
to the classical action. Therefore, we need more. Since we do not have free parameters to play with (an overall inverse temperature
factor in front of the GFT action can be reabsorbed with a field redefinition), we have to conjecture that the semiclassical
limit will require some other large number to appear in the action. Again, for naturalness reasons, this number can be associated
only to the presence of a large number of additional degrees of freedom, \ie to a large number of matter fields. 
Obviously this conjecture must be carefully proven, but it would match some rather old ideas on the semiclassical
limit of quantum gravity \cite{hartlehorowitz}.

We want to conclude with one of the most puzzling points of this paper.
In section \ref{models} we have said that the models that we are playing with are leading to equilateral
simplices, and that all the information about curvature is lost. Of course, this is a very unpleasant situation,
at a first glance. However, the results of sections \ref{WDW2} and \ref{characteristiccurve} clearly show that,
despite this trivial short range dynamics, the long range dynamics, controlled by the properties of the critical point,
might be nonetheless highly nontrivial, at the point of resembling the symmetry reduced dynamics of gravitational theories. 

This might be expected in view of the general argument of universality \cite{cardybook,goldenfeldbook}:
many different microscopic Hamiltonians, belonging to the same basin of attraction of a given critical point,
will lead to indistinguishable critical behavior. This might be seen as an hint that perhaps, at this level of refinement,
it is enough to construct a model with the correct universality class, rather than with the correct microscopic dynamics.

This of course is still a difficult task, as the understanding of dynamical triangulations versus causal dynamical triangulations has told us. There, the inclusion at the microscopic level of all the relevant terms (the discretized Einstein--Hilbert action with cosmological constant) does not guarantee that the continuum limit will describe four dimensional spacetimes. Therefore,
it might be well that the implementation of a microscopic GFT model might be complicated by the fact that the implementation
of a simplicial path integral for gravity might be less important than the identification of the correct universality class, which is ultimately controlling the macroscopic limit. Therefore, the classification of GFT universality classes stands out as one of the most important problems that such a formalism has to face. 

On this possible mismatch between microscopic and microscopic level, the analysis presented in \cite{oritisindoni} about the effective actions around prescribed GFT vacua shows
that, if these effective dynamics encodes the gravitons, their action might be quite different from the fundamental action
with which we start building spacetime from nothing. This of course complicates the path towards a formulation of concrete recipes that can tell us how to build the microscopic theory once we know what is the macroscopic limit we want to approach.

As mentioned in the introduction, this might be a coincidence. In fact, we can argue for the contrary. After all, the system
we are playing with is still a system in which we are trying to realize a phase transition that will lead to some notion of
continuum geometry. By taking the point of view of Landau, whatever is the microscopic theory, the macroscopic order
parameter we should expect is a three geometry (in the case of the wavefunction). The appearance of an effective
geometrodynamics will be just a consequence of the fact that this order parameter will have to satisfy some equations of motion
(Ginzburg--Landau equations), which are some form of geometrodynamics. Whether these equations are leading to
general relativity or to some other theory is a more delicate matter, that will be carefully examined in future work.

\section*{Acknowledgments}
I would like to thank Daniele Oriti and Stefano Liberati, for constructive criticism on previous version of the draft, and for inspiring discussions.

\end{document}